\documentstyle[aps,preprint,floats,tighten]{revtex}
\input psfig
\begin{document}
\draft
\preprint{\vbox{\hbox{} 
                \hbox{}
                \hbox{}
}}

\title{A Polarization Pursuers' Guide}

\author{Andrew H. Jaffe$^1$\footnote{jaffe@cfpa.berkeley.edu},
Marc Kamionkowski$^{2,3}$\footnote{Email: 
kamion@tapir.caltech.edu}, and Limin Wang$^2$\footnote{Email:
limin@cuphyb.phys.columbia.edu}}
\address{$^1$Center for Particle Astrophysics, 301 LeConte Hall,
University of California, Berkeley, CA 94720}
\address{$^2$Department of Physics, 538 West 120th Street,
Columbia University, New York, NY 10027}
\address{$^3$California Institute of Technology, Mail Code
130-33, Pasadena, CA 91125}
\date{September 1999}
\maketitle

\begin{abstract}
We calculate the detectability of the polarization of the
cosmic microwave background (CMB) as a function of the sky
coverage, angular resolution, and instrumental sensitivity for a
hypothetical experiment.  We consider the gradient component of
the polarization from density perturbations (scalar modes) and
the curl component from gravitational waves (tensor modes).  We
show that the amplitude (and thus the detectability) of the
polarization from density perturbations is roughly the same in
any model as long as the model fits the big-bang-nucleosynthesis
(BBN) baryon density and degree-scale anisotropy measurements.
The degree-scale polarization is smaller (and accordingly more
difficult to detect) if the baryon density is higher.  
In some cases, the signal-to-noise for polarization (both from
scalar and tensor modes) may be improved in a fixed-time
experiment with a smaller survey area.
\end{abstract}


\def\hatn{{\bf \hat n}}
\def\hatnprime{{\bf \hat n'}}
\def\hatnone{{\bf \hat n}_1}
\def\hatntwo{{\bf \hat n}_2}
\def\hatni{{\bf \hat n}_i}
\def\hatnj{{\bf \hat n}_j}
\def\vecx{{\bf x}}
\def\veck{{\bf k}}
\def\hatx{{\bf \hat x}}
\def\hatk{{\bf \hat k}}
\def\hatz{{\bf \hat z}}
\def\VEV#1{{\left\langle #1 \right\rangle}}
\def\cP{{\cal P}}
\def\noise{{\rm noise}}
\def\pix{{\rm pix}}
\def\map{{\rm map}}
\long\def\comment#1{}
\newcommand{\be}{\begin{equation}}
\newcommand{\ee}{\end{equation}}
\newcommand{\bea}{\begin{eqnarray}}
\newcommand{\eea}{\end{eqnarray}}

\section{Introduction}

It has long been known that the cosmic microwave background
(CMB) must be polarized if it has a cosmological origin \cite{rees,arthur}.
Detection, and ultimately mapping, of the polarization will
help isolate the peculiar velocity at the surface of last
scatter \cite{ZalHar95}, constrain the ionization history of
the Universe \cite{Zal97}, determine the nature of primordial
perturbations \cite{Kos98,SpeZal97}, detect an inflationary
gravitational-wave background \cite{ourletter,szletter},
primordial magnetic fields
\cite{KosLoe96,HarHayZal96,ScaFer97}, and cosmological parity
violation \cite{LueWanKam98,Lep98}, and maybe more (see, e.g.,
Ref. \cite{KamKos99} for a recent review).  However, the
precise amplitude and angular spectrum of the polarization
depends on a structure-formation model and the values of
numerous undetermined parameters.  Moreover, it has so far
eluded detection.

A variety of experiments are now poised to detect
the polarization for the first time.  But what is the ideal
experiment?  What angular resolution, instrumental sensitivity,
and fraction of the sky should be targeted?  Can it be picked
out more easily by cross-correlating with the CMB temperature?
The purpose of this paper is to answer these questions in a 
fairly model-independent way.

We first address the detectability of the gradient component of
the polarization from density perturbations.
{\it A priori}, one might expect the detectability of this
signal to depend sensitively on details of the
structure-formation model, ionization history, and on a variety
of undetermined cosmological parameters.  However, we find that
if we fix the baryon-to-photon ratio to its
big-bang-nucleosynthesis (BBN) value and demand that the
degree-scale anisotropy agree with recent measurements, then the 
detectability of the polarization is roughly model-independent.
We provide some analytic arguments in support of this result.
We can thus specify an experiment that would
be more-or-less guaranteed to detect the CMB polarization.
Non-detection in such experiments would thus only be explained
if the baryon density considerably exceeded the BBN value.

We then consider the curl component of the polarization
{}from an inflationary gravitational-wave background.  This
extends slightly the work of
Refs. \cite{marcarthur,Lesetal99}.\footnote{There is also
related work in Refs. \cite{Kin98,ZalSelSpe97,Lidetal97b} in
which it is determined how accurately various cosmological and
inflationary parameters can be determined in case of a positive
detection.}  The new twist here is that we consider maps with
partial sky coverage (rather than only full-sky maps) and find
that in a noise-limited fixed-time experiment, the sensitivity
to gravitational waves may be improved considerably by surveying
more deeply a smaller patch of sky.\footnote{Similar arguments
were investigated for temperature maps in
Refs. \cite{MagHob97,HobMag96}.}

Since the polarization should be detected shortly, our main
results on its detectability should, strictly speaking, become
obsolete fairly quickly.  Even so, our results should
be of some lasting value as they provide figures of merit for
comparing the relative value, in
terms of signal-to-noise, of various future CMB polarization
experiments.  It should be kept in mind, however, that ours is a
hypothetical experiment in which foregrounds have been
subtracted and instrumental artifacts understood, and any
comparison with realistic experiments must take these effects
into account.

Section II briefly reviews the CMB polarization signals.  Section III
introduces the formalism for determining the detectability of
polarization for a given experiment.  Section IV considers
polarization from scalar modes for a putative structure-formation model
and Section V evaluates the detectability of the polarization from
gravitational waves (using only the curl component of the polarization).
Section VI presents the results of the prior two Sections in a slightly
different way.  Section VII shows that the results for scalar
modes in Section IV
would be essentially the same in virtually any other structure-formation
model with a BBN baryon density and degree-scale temperature anisotropy
that matches recent measurements.  We make some concluding
remarks in Section VII.

\section{Brief Review of CMB Polarization}

Ultimately, the primary goal of CMB polarization experiments
will be to reconstruct the polarization power spectra and the
temperature-polarization power spectrum.  Just as a temperature
(T) map can be expanded in terms of spherical harmonics, a
polarization map can be expanded in terms of a set of tensor
spherical harmonics for the gradient (G) component of the
polarization and another set of harmonics for the curl (C)
component \cite{ourlongpaper,szlongpaper}.
Thus, the two-point statistics of the T/P map are specified
completely by the six power spectra $C_\ell^{{\rm X}{\rm X}'}$
for ${\rm X},{\rm X}' = \{{\rm T,G,C}\}$.  Parity invariance
demands that $C_\ell^{\rm TC}=C_\ell^{\rm GC}=0$ (unless the physics
that gives rise to CMB fluctuations is parity breaking
\cite{LueWanKam98}).  Therefore the statistics of the CMB
temperature-polarization map are completely specified by the
four sets of moments, $C_\ell^{\rm TT}$, $C_\ell^{\rm TG}$, $C_\ell^{\rm
GG}$, and $C_\ell^{\rm CC}$.  See, e.g., Fig.~1 in
Ref. \cite{marcarthur} for sample spectra from adiabatic
perturbations and from gravitational waves.  

There are essentially two things we would like to do with
the CMB polarization: (1) map the G component to study
primordial density perturbations, and (2) search for the
C component due to inflationary gravitational waves
\cite{ourletter,szletter}.  The G
signal from density perturbations is more-or-less guaranteed to
be there at some level (to be quantified further below), and
will undoubtedly provide a wealth of information
on the origin of structure and the values of cosmological
parameters.  The amplitude of the C component from
inflationary gravitational waves is proportional to the
square of the (to-be-determined) energy scale of inflation.  It
is not guaranteed to be large enough to be detectable even if
inflation did occur.  On the other hand, if inflation had
something to do with grand unification or Planck-scale physics,
as many theorists surmise, then the polarization is conceivably
detectable, as argued in Refs. \cite{KamKos99,marcarthur,Kin98}
and further below.  If detected, it would provide a unique and
direct window to the Universe as it was $10^{-36}$ seconds after
the big bang!

\section{Formalism}

We first address the general question of the detectability of a particular
polarization component.  We assume that the amplitude of the various
polarization signals will each be picked out by a maximum-likelihood
analysis \cite{bjk98}.  The shape of the likelihood function will then
give limits on the parameters, in this case the various power spectra,
$C_\ell^{\rm XX}$ that describe the CMB. In particular, the curvature of the
likelihood gives traditional error bars, defined as for a Gaussian
distribution (but see Ref. \cite{bjk99} for a discussion of the more
complicated true non-Gaussian distribution). Here we will concentrate on 
the error bar for the overall amplitude of the power spectra.

We can then ask, what is the smallest amplitude that could be
distinguished from the null hypothesis of no polarization component by
an experiment that maps the polarization over some fraction of the sky
with a given angular resolution and instrumental noise?  This question
was addressed (for the curl component) in Ref. \cite{marcarthur}
for a full-sky map.  If an
experiment concentrates on a smaller region of sky, then several things
happen that affect the sensitivity: (1) information from modes with
$\ell\lesssim180/\theta$ (where $\theta^2$ is the area on the sky mapped)
is lost;\footnote{This is not strictly true. In principle, as usual in
  Fourier analysis, less sky coverage merely limits the independent
  modes one can measure to have a spacing of $\delta
  l\gtrsim180/\theta$. In practice, instrumental effects (detector
  drifts; ``1/f'' noise) will render the smallest of these bins
  unobservable.}  (2) the sample variance is increased; (3) the noise
per pixel is decreased since more time can be spent integrating on this
smaller patch of the sky.

For definiteness, suppose we hypothesize that there is a C component of
the polarization with a power spectrum that has the $\ell$ dependence
expected from inflation (as shown in Fig.~1 in Ref. \cite{marcarthur}),
but an unknown amplitude ${\cal T}$.\footnote{We define ${\cal T}\equiv
  6 C_2^{\rm TT, tens}$ where $C_2^{\rm TT, tens}$ is the tensor
  contribution to the temperature quadrupole moment expected for a
  scale-invariant spectrum.}

We can predict the size of the error that we will obtain from the
ensemble average of the curvature of the likelihood function (also known 
as the Fisher matrix) \cite{jungmanone,jungmantwo}. For example,
consider the tensor signal.  In such a likelihood analysis, the
expected error will be $\sigma_{\cal T}$, where
\begin{equation}
     {1\over \sigma_{\cal T}^2} = \sum_\ell \left( { \partial C_\ell^{\rm CC}
     \over \partial {\cal T}} \right)^2 {1\over
     (\sigma_\ell^{\rm CC})^2},
\label{simplest}
\end{equation}
with similar equations for the other $C_\ell^{\rm XX}$. 

Here, the $\sigma_\ell^{{\rm XX}'}$ are the expected errors at
individual $\ell$ for each of the ${\rm XX}'$ power
spectra. These are given by (cf., Ref. \cite{ourlongpaper})
\begin{eqnarray}
  \label{eq:sigl}
  \sigma^{\rm CC}_\ell &=& \sqrt{2\over f_{\rm sky}(2\ell+1)}
  \left(C_\ell^{\rm CC} + f_{\rm sky} w^{-1} B_\ell^{-2}\right),
\nonumber\\ 
  \sigma^{\rm GG}_\ell &=& \sqrt{2\over f_{\rm sky}(2\ell+1)}
  \left(C_\ell^{\rm GG} + f_{\rm sky} w^{-1} B_\ell^{-2}\right),
\nonumber\\ 
  \sigma^{\rm TG}_\ell &=& \sqrt{1\over f_{\rm sky}(2\ell+1)}
  \left[\left(C_\ell^{\rm TG}\right)^2 + 
    \left(C_\ell^{\rm TT}+f_{\rm sky} w^{-1} B_\ell^{-2}\right)
    \left(C_\ell^{\rm GG}+f_{\rm sky} w^{-1} B_\ell^{-2}\right)  \right]^{1/2},
\end{eqnarray}
where $w=(t_{\rm pix} N_{\rm pix}T_0^2)/(4\pi s^2)$ is the weight
(inverse variance) on the sky spread over $4\pi$ steradians, $f_{\rm
  sky}$ is the fraction of the sky observed, and $t_{\rm pix}$ is the
time spent observing each of the $N_{\rm pix}$ pixels.
The detector sensitivity is $s$ and the average sky temperature
is $T_0=2.73\,\mu{\rm K}$ (and hence all $C_\ell^{\rm XX'}$ are measured
in dimensionless $\Delta T/T$ units). The inverse weight for a full-sky
observation is $w^{-1}=2.14\times10^{-15} t_{\rm yr}^{-1}(s/200\,
\mu{\rm K}\, \sqrt{\rm sec})^2$ with $t_{\rm yr}$ the total observing
time in years. Finally, $B_\ell$ is the experimental beam, which for a
Gaussian is $B_\ell=e^{-\ell^2 \sigma_\theta^2/2}$.  We assume all
detectors are polarized.  As mentioned, all other $C_\ell^{{\rm XX}'}$
cross terms are zero (in the usual cases, at least).

The CC and GG errors each have two terms, one proportional to
$C_\ell^{\rm XX}$ (the {\em sample variance}), and another proportional to
$w^{-1}$ (the {\em noise variance}). The TG error is more complicated
since it involves the product of two different fields (T and G)
on the sky.

There are several complications to note when considering these formulae:
1) We never have access to the actual $C_\ell^{\rm XX'}$, but only to
some estimate of the spectra; 2) the expressions only deal approximately
with the effect of partial sky coverage; and 3) the actual likelihood
function can be considerably non-Gaussian, so the expressions above do
not really refer to ``1 sigma confidence limits.''

Here, we are interested in the detectability of a polarization
component; that is, what is the smallest polarization amplitude that we
could confidently differentiate from zero? This answer in detail depends
on the full shape of the likelihood function: the ``number of sigma''
that the likelihood maximum lies away from zero is related to the
fraction of integrated likelihood between zero and the maximum.  This
gives an indication of how well the observation can be distinguished
{}from zero power in the polarization.  Toy problems and experience give
us an approximate rule of thumb: the signal is detectable when it can be
differentiated from the ``null hypothesis'' of $C_\ell^{\rm XX}=0$.
Stated another (more Bayesian) way, for a fixed noise variance, as we
increase the observed signal the fraction of probability below the peak
increases rapidly when the sample variance---i.e., the estimated
power---approaches the noise.  Thus, on the one hand you need to observe
enough sky to sufficiently decrease the sample variance, and a small
enough noise that the sample variance dominates.

Thus, the $\ell$ component of the tensor signal (for example) is
detectable if its amplitude is greater than
\begin{equation}
     \sigma_\ell^{\rm CC} = \sqrt{2/(2\ell+1)} f_{\rm sky}^{1/2} 
     w^{-1} e^{\ell^2 \sigma_b^2}.
\label{CCnoise}
\end{equation}
We then estimate the smallest tensor amplitude ${\cal T}$ that
can be distinguished from zero (at ``1 sigma'') by using
Eq. (\ref{simplest}) with the null hypothesis $C_\ell^{\rm
TT}=0$.  Putting it all together, the smallest detectable tensor
amplitude (scaled by the largest consistent with COBE) is
\begin{equation}
     {\sigma_{\cal T} \over {\cal T}} \simeq
     1.47\times10^{-17}\, t_{\rm yr} \, \left( {s \over 200\,
     \mu {\rm K}\sqrt{\rm sec}} \right)^2 \, \left({\theta \over 
     {\rm deg}} \right) \, \Sigma_\theta^{-1/2},
\end{equation}
where
\begin{equation}
     \Sigma_\theta = \sum_{\ell\geq (180/\theta)} (2\ell+1) \left(
     C_\ell^{\rm CC} \right)^2 e^{-2\ell^2\sigma_b^2}.
\label{summand}
\end{equation}
The expression for the GG signal from density perturbations is
obtained by replacing $C_\ell^{\rm CC}$ by $C_\ell^{\rm GG}$ [and
$\sigma_\ell^{\rm GG}$ is the same as $\sigma_\ell^{\rm CC}$ given in
Eq. (\ref{CCnoise})].

For the TG cross-correlation, things are more complicated. First of all, 
the expression for $\sigma_\ell^{\rm TG}$ in Eq.~\ref{eq:sigl} has terms
involving the temperature power spectra and observing
characteristics. Second, we {\em know} that there is a temperature
component on the sky, so we must pick a ``null hypothesis'' with the
observed $C_\ell^{\rm TT}$.  The TG moments also have
covariances with the TT and GG moments (see Eqs. (3.28)--(3.30)
in Ref. \cite{ourlongpaper}), but these are zero for the null
hypothesis of no polarization.  Hence, with the null hypothesis
of no polarization, the variance with which each $C_\ell^{\rm
TG}$ can be measured is
\begin{equation}
     \sigma_\ell^{\rm TG} = \sqrt{1/f_{\rm sky}(2\ell+1)} \left[
     f_{\rm sky} w^{-1} e^{\ell^2 \sigma_b^2} \left(C_\ell^{\rm TT} +
     f_{\rm sky} w^{-1} e^{\ell^2 \sigma_b^2} \right) \right]^{1/2}.
\label{TGnoise}
\end{equation}
Thus, the dependence on $s$ is more complicated than in
Eq. (\ref{TGnoise}), so the end result for the polarization
sensitivity achievable by cross-correlating with the temperature 
does not scale simply with $s$ or $t_{\rm yr}$ as it does for GG
and CC.

\section{Detectability of Density-Perturbation Signal}
\label{dpsubsec}

Since the polarization has yet to be detected, the obvious first 
goal of a current experiment should be to detect unambiguously the
polarization. In the standard theory with adiabatic perturbations
somehow laid down prior to last scattering, the G polarization is
inevitable.
Density perturbations will thus produce a nonzero GG power spectrum
and a TG power spectrum.  We discuss the detectability of
polarization from only the GG signal or the TG signal
individually, and then from the combination of both signals.

\begin{figure}[htbp]
\centerline{\psfig{file=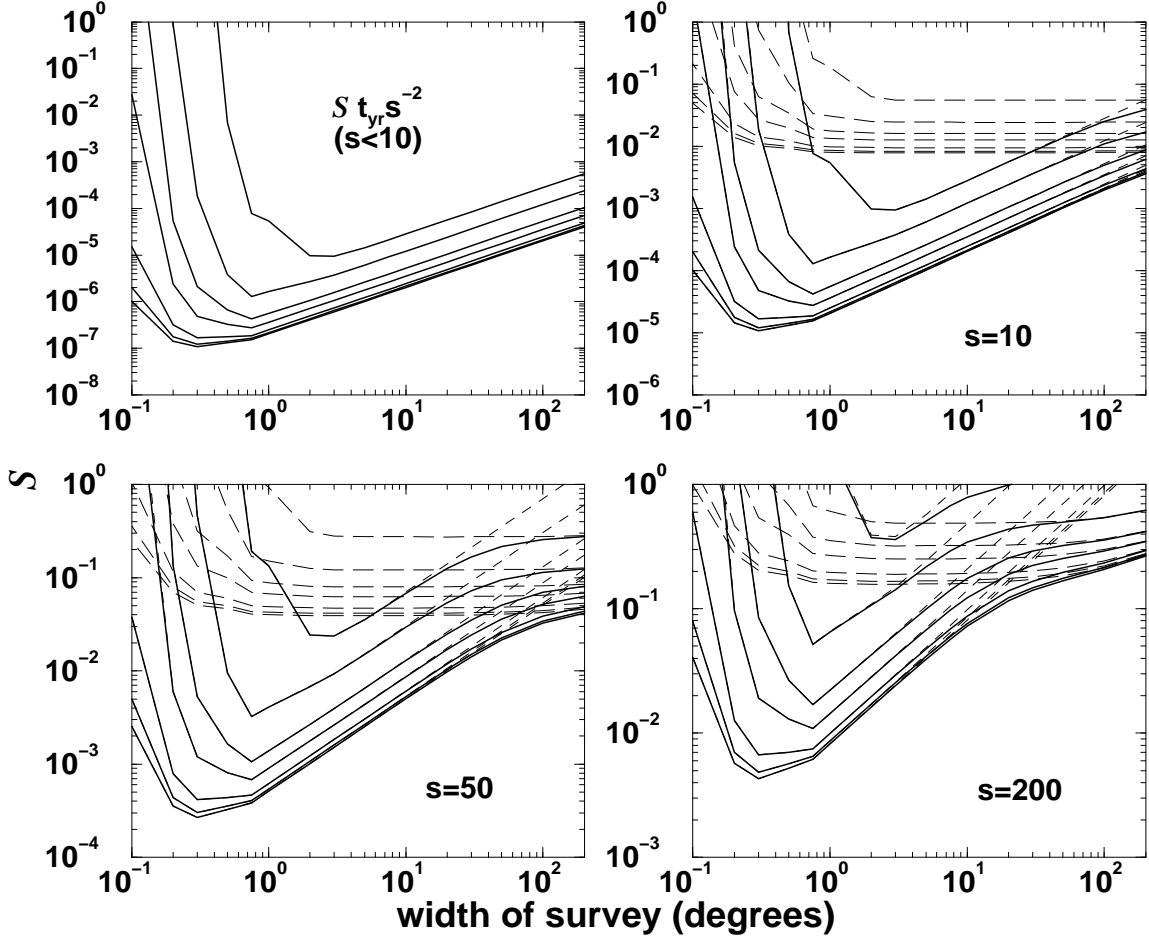,width=6in}}
\bigskip
\caption{The smallest amplitude ${\cal S}$ of the polarization
         signal from density perturbations (scaled by that
         expected from a COBE-normalized CDM model)
         that could be detected at $3\sigma$ with various values
         of the detector sensitivity $s$ (given in units of $\mu
         {\rm K} \sqrt{\rm sec}$) for an experiment that runs
         for one year and maps a square region of the sky of a
         given width.  The long-dash curves show the
         sensitivities achievable by cross-correlating with the
         CMB temperature.  The short-dash curves show
         sensitivities achievable using only the polarization
         autocorrelation function (the GG power spectrum).  The
         solid curves show results achievable using both the GG
         and TG power spectra.  The curves are (from top to
         bottom) for fwhm beamwidths of 1, 0.5, 0.3, 0.2, 0.1,
         0.05, and 0.01 degrees.  For $s< 10\,\mu {\rm K}
         \sqrt{\rm sec}$, the sensitivity comes entirely from
         the polarization auto-correlation and scales as $s^2$
         and inversely with $t_{\rm yr}$, as shown in the top
         left-hand panel.}
\label{fig:scalarsens}
\end{figure}

\subsection{The GG signal}

We have calculated the detectability of the polarization from density
perturbations (using only the GG power spectrum), and results are shown
in Fig.~\ref{fig:scalarsens}.  Here, we ask the following: Suppose there
is a polarization signal with an $\ell$ dependence characteristic of
density perturbations but of unknown amplitude.  What is the smallest
polarization amplitude that could be distinguished from the null
hypothesis of no polarization?  The short-dashed curves (which coincide
with the solid curves for small survey widths) in
Fig.~\ref{fig:scalarsens} show the smallest polarization amplitude
${\cal S}$ (scaled by that expected for a COBE-normalized CDM model)
detectable (at $3\sigma$) by an experiment with detector sensitivity
$s$ that maps the polarization only on a
square region of the sky with a given width.  The curves are (from top
to bottom) for fwhm beamwidths of 1, 0.5, 0.3, 0.2, 0.1, 0.05, and
0.01 degrees.  The results scale with the square of the detector
sensitivity and inversely to the duration of the experiment. 
Any experiment that has a $(\sigma_{\cal S}/{\cal S})$ smaller
than unity should be able to detect the polarization expected in
a CDM model at $>3\sigma$.  
Fig. \ref{fig:scalarsens} shows that
an experiment with comparable $s$ can in a few months achieve
the same signal-to-noise with the GG power spectrum as MAP.

\subsection{The TG signal}

We have also done the same analysis for the
temperature/polarization cross correlation (the TG power
spectrum), and the results are indicated by the long-dashed
curves in Fig.~\ref{fig:scalarsens}.  Here we ask,
what is the smallest polarization signal that could be
distinguished from the null hypothesis of no polarization
by looking for the expected temperature-polarization
cross-correlation?  First of all, the analog of Eq. (\ref{CCnoise})
for GG is given for TG by Eq. (\ref{TGnoise}).  Thus,
cosmic variance in the temperature map comes into play even if
we investigate the null
hypothesis of no polarization.  As a result, the detectability
of the polarization from temperature-polarization
cross-correlation does not scale simply with the instrumental
sensitivity $s$ (and this is why we present results for
detectability in four panels for four different values of $s$ in
Fig.~\ref{fig:scalarsens} rather than on one panel as we will for 
the curl component in Fig.~\ref{fig:tensorsens}). However,
comparing the long-dash curves in all four panels, we see that
for $s\lesssim200\,\mu {\rm K} \sqrt{\rm sec}$ (reasonable
values for just about any future experiments), the
detector-noise term in Eq. (\ref{TGnoise}) is less important
than the $C_\ell^{\rm TT}$ term, so the result for $\sigma_{\cal
S}/{\cal S}$ scales with $s$.

\subsection{Combining the TG and GG Signal}

Comparing the long- and short-dash curves in
Fig.~\ref{fig:scalarsens}, we see that the polarization
sensitivity obtained by looking for a temperature-polarization
cross-correlation improves on that obtained from the
polarization auto-correlation (for fixed angular resolution and
detector sensitivity) only for nearly full-sky surveys with
$s\gtrsim50 \, \mu {\rm K} \sqrt{\rm sec}$.  Thus, the
sensitivity of MAP (full-sky and $s\simeq 150 \mu {\rm K}
\sqrt{\rm sec}$) to polarization will come primarily from
cross-correlating with the temperature map, while the
signal-to-noise for polarization auto-correlation and
temperature-polarization cross-correlation should be roughly
comparable for Planck.  The Figure also indicates that
in an experiment with $s\lesssim 100 \mu {\rm K} \sqrt{\rm sec}$ 
that maps only a small fraction of the sky (widths $\lesssim
10^\circ$), the polarization is more easily detected via
polarization auto-correlations; cross-correlating with the
temperature should not significantly improve the prospects for
detecting the polarization in such experiments.

The total sensitivity achievable using both TG and GG together is
obtained by adding the sensitivities from each in
quadrature.\footnote{In principle, there are cross terms between TG and
  GG in the correlation matrix.  However, for the null hypothesis of no
  polarization, these are zero; cf., Eq. (3.29) in Ref.
  \cite{ourlongpaper}.}  The solid curves in Fig.~\ref{fig:scalarsens}
show the polarization sensitivities achievable by combining the GG and
TG data.  For $s\lesssim 10\,\mu {\rm K} \sqrt{\rm sec}$, the
sensitivity comes entirely from the polarization auto-correlation and
scales as $s^2$, as shown in the top left-hand panel.

We see that the sensitivity to the polarization can improve as
the angular resolution is improved all the way down to 0.01
degrees, and the ideal survey size varies from 2--3 degrees (for
an angular resolution of 1 degree) to a fraction of a degree for
better angular resolution.

\section{Detectability of the Curl Component}

Consider next the C component, which can tell us about the amplitude of
gravitational waves produced, for example, by inflation. We have
carried out the
exercise as for the scalar signal. As above we hypothesize that there
is a C component of the polarization with an unknown amplitude ${\cal T}$

\begin{figure}[htbp]
\centerline{\psfig{file=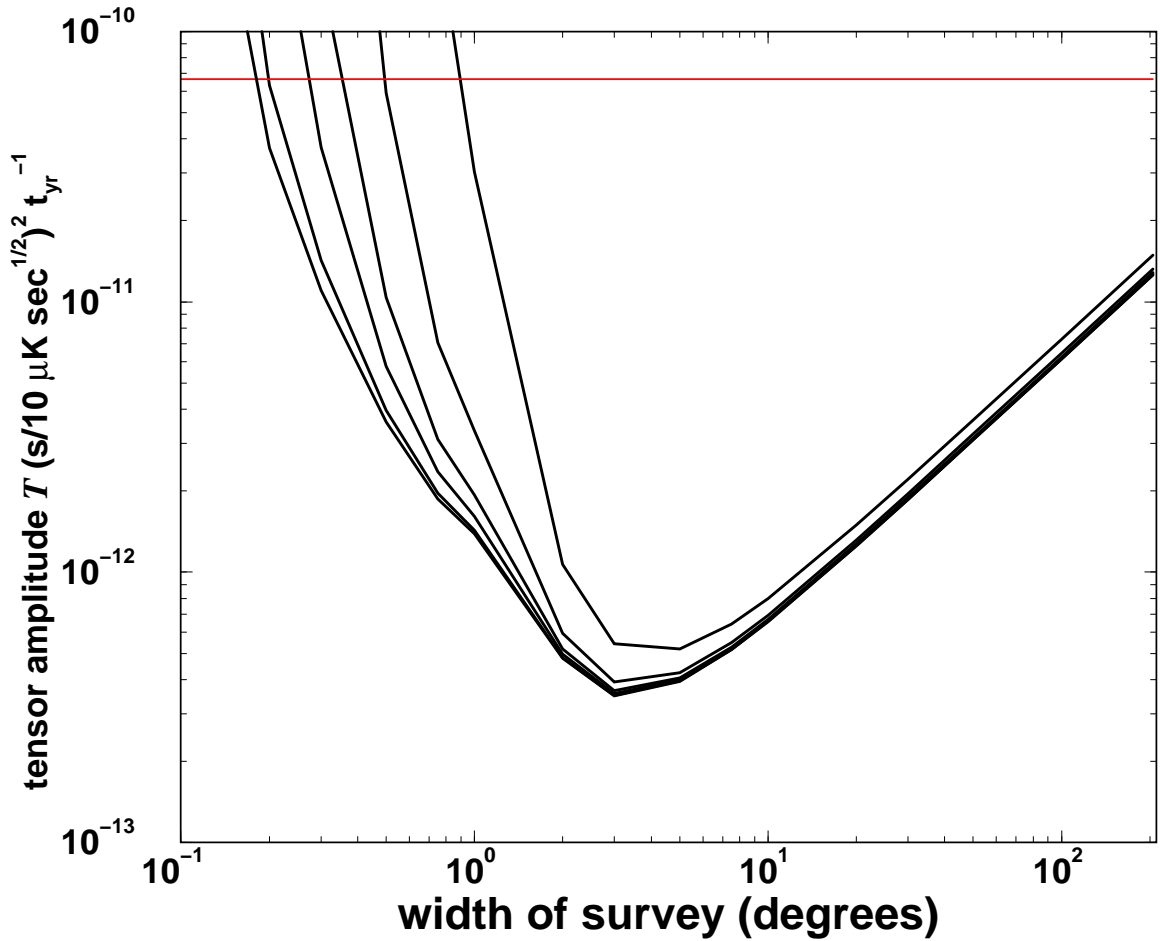,width=6in}}
\bigskip
\caption{The smallest tensor amplitude ${\cal T}$ that
         could be detected at $3\sigma$ with an experiment with a detector
         sensitivity $s = 10\,\mu {\rm K} \sqrt{\rm sec}$ that
         runs for one year and maps a square region of the sky
         of a given width.  The result scales with the square of 
         the detector sensitivity and inversely with the
         duration of the experiment.  The curves are (from top
         to bottom) for fwhm beamwidths of 1, 0.5, 0.3, 0.2,
         0.1, and 0.05 degrees.  The horizontal line shows the upper limit 
         to the tensor amplitude from COBE.}
\label{fig:tensorsens}
\end{figure}

Results are shown in Fig.~\ref{fig:tensorsens}.  Plotted there is the
smallest gravitational-wave (i.e., tensor) amplitude ${\cal T}$
detectable at $3\sigma$ by an experiment with a detector sensitivity
$s=10\,\mu {\rm K} \sqrt{\rm sec}$ that maps a square region of the sky
over a year with a given beamwidth.  The horizontal line shows the upper
limit to the tensor amplitude from COBE.  The curves are (from top to
bottom) for fwhm beamwidths of 1, 0.5, 0.3, 0.2, 0.1, and 0.05 degrees.
The results scale with the square of the detector sensitivity and
inversely to the duration of the experiment.

The sensitivity to the tensor signal is a little better
with an 0.5-degree beam than with a 1-degree beam, but even
smaller angular resolution does not improve the sensitivity
much.  And with a resolution of 0.5 degrees or better, the best
survey size for detecting this tensor signal is about 3 to 5
degrees.  If such a fraction of the sky is surveyed, the
sensitivity to a tensor signal (rms) will be about 30 times
better than with a full-sky survey with the same detector
sensitivity and duration (and thus 30 times better than
indicated in Refs. \cite{KamKos99,marcarthur}.  Thus, a balloon
experiment with the same detector sensitivity as MAP could in
principle detect the same tensor amplitude in a few weeks that
MAP would in a year.  (A width of 200 degrees corresponds to
full-sky coverage.)

The tensor amplitude is related to the energy scale of
inflation\footnote{The energy scale of inflation is defined here
to be the fourth root of the
inflaton-potential height.} by ${\cal T}=(E_{\rm
infl}/7\times10^{18}\, {\rm GeV})^4$, and COBE currently
constrains $E_{\rm infl} \lesssim 2\times10^{16}$ GeV
\cite{KamKos99,ZibScoWhi99}.  Thus with
Fig.~\ref{fig:tensorsens}, one can determine the inflationary
energy scale accessible with any given experiment.

\section{SOME DETAILS AND INSIGHT}

\begin{figure}[htbp]
  \begin{center}
    \psfig{file=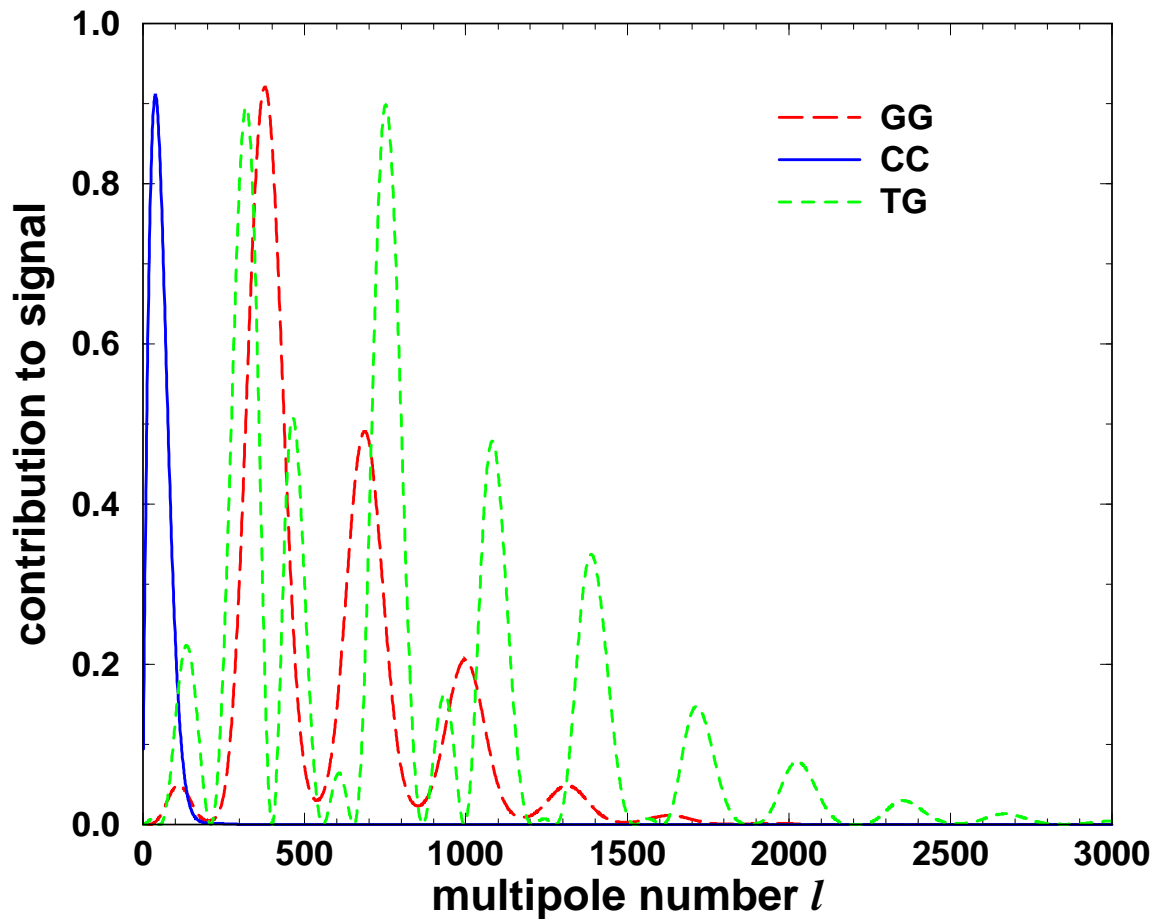,width=6in}
  \end{center}
  \caption{
The summands for $\Sigma_\theta$ for the CC, TG, and GG signals
      for a full-sky map with perfect angular resolution and no detector
      noise.  The signal-to-noise of each signal is proportional to the
      inverse of the square root of the area under the curve.  A beam of
      finite width $\theta_{\rm fwhm}$ would reduce the contribution
      {}from $\ell{>} 200(\theta_{\rm fwhm}/{\rm deg})^{-1}$'s and a
      survey of width $\theta$ would reduce the contribution from
      $\ell{<}(\theta/180\, {\rm deg})$'s.
}
\label{fig:conts}
\end{figure}

Fig.~\ref{fig:conts} is intended to provide some additional insight into
the results shown in Figs.~\ref{fig:scalarsens} and
\ref{fig:tensorsens}.  Fig.~\ref{fig:conts} plots the summands (with
arbitrary normalization) from Eq. (\ref{summand}) for $\Sigma_\theta$
for the CC and GG signals for a full-sky map with perfect angular
resolution.  It also shows the analogous summand for TG.  The
detectability of each signal (CC, GG, and TG) is inversely proportional
to the square root of the area under each curve.  A finite beamwidth
(and/or instrumental noise) would reduce the contribution from higher
$\ell$'s and a survey area less than full-sky would reduce the contribution
{}from lower $\ell$'s.  The Figure illustrates that the CC signal is best
detected with $\ell\lesssim200$ and the GG signal is best detected with
$\ell\simeq200-1200$, as may have been surmised from Figs.
\ref{fig:scalarsens} and \ref{fig:tensorsens}.  The TG signal is spread
over a larger range of $\ell$'s.  In particular, note that very little of
$\Sigma_\theta$ comes from $\ell\lesssim10$ in any case, so the loss of the
$\ell\lesssim10$ modes that comes with survey regions smaller than
$10\times10$ deg$^2$ does not significantly affect the detectability of
the polarization signals.

And now for some historical perspective.  Although not shown,
the $\Sigma_\theta$ for the temperature power spectrum TT
peaks {\it very} sharply at low $\ell$ (it essentially falls off as
$\ell^{-3}$ for a nearly scale-invariant spectrum
[$\ell(\ell+1)C_\ell\simeq{\rm const}$] such as that
observed).  Thus, in retrospect, the COBE full-sky scan was
indeed the best strategy for detecting the temperature
anisotropy.  An equal-time survey of a smaller region of the sky 
would have made detection far less likely.

\section{Model Independence of the Results}

In Section \ref{dpsubsec} we used a standard-CDM model for our
calculations, and it is natural to inquire whether and/or how
our results depend on this assumption.  The purpose of this
Section is to illustrate that the results shown above are, to a
large extent, {\it in}dependent of the gross features and
details of the structure-formation model as long as we (1) use
the BBN baryon density and (2) demand that the model reproduce
the degree-scale anisotropy observed by several recent
experiments.

\begin{figure}[htbp]
\centerline{\psfig{file=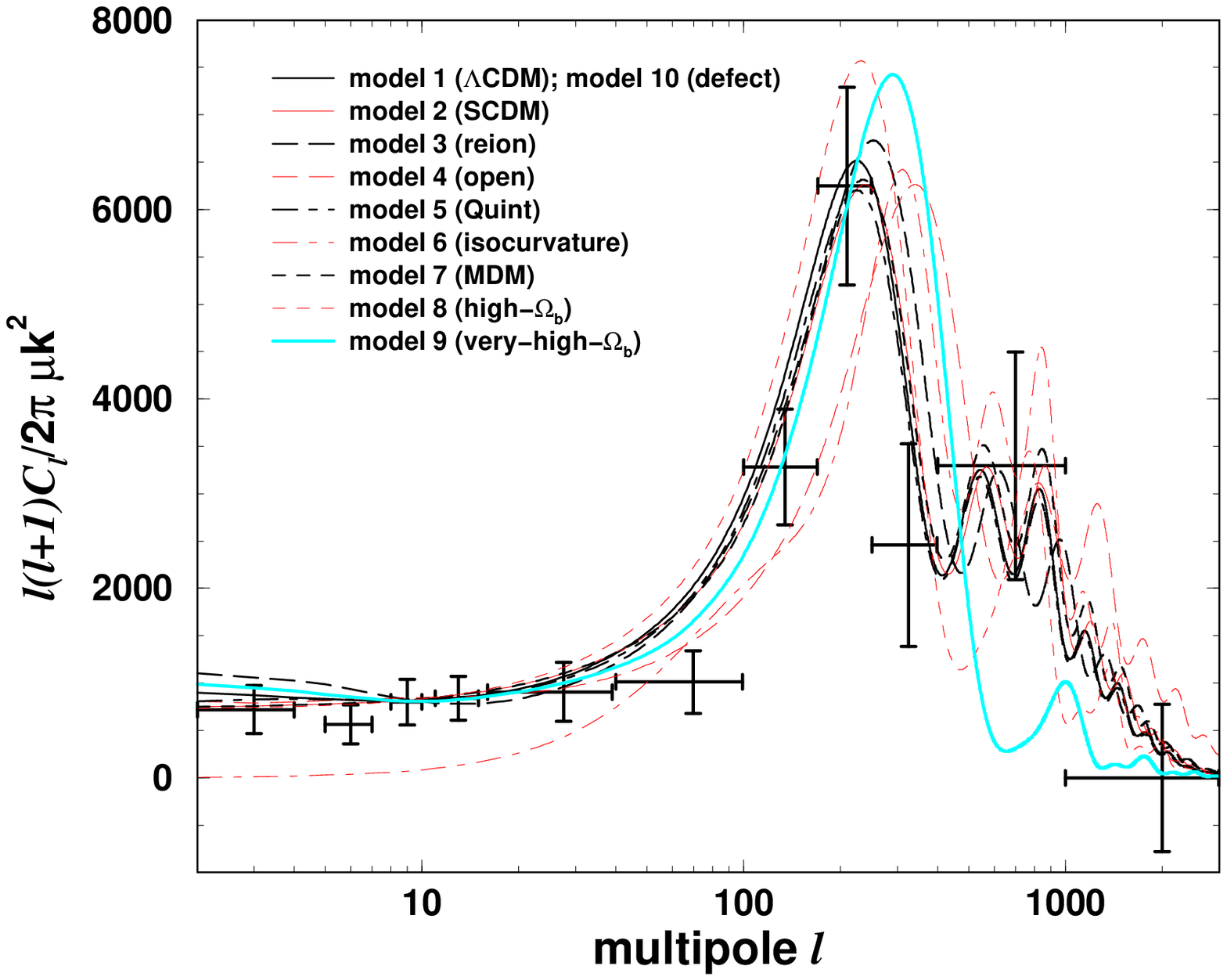,width=6in}}
\bigskip
\caption{The temperature power spectra for the
         structure-formation models listed in Table \ref{table}.  
         The models were all chosen to fit (by eye) the
         data points (see Ref. \protect\cite{bjk99} for how the data points 
	 were compiled) near $\ell\sim200$.}
\label{fig:CTls}
\end{figure}

\begin{figure}[htbp]
\centerline{\psfig{file=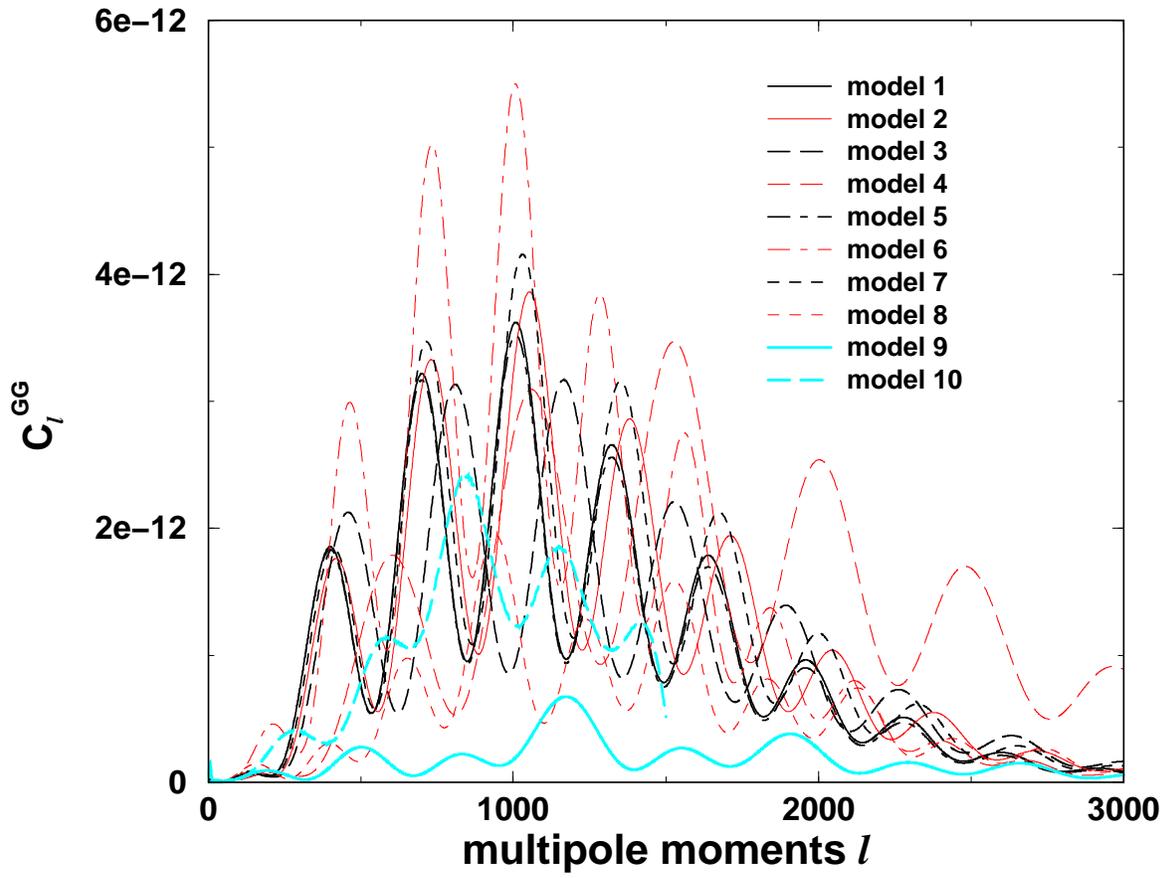,width=6in}}
\bigskip
\caption{The GG polarization power spectra for each of the
         models listed in Table \ref{table}.}
\label{fig:ClGG}
\end{figure}

\begin{figure}[htbp]
\centerline{\psfig{file=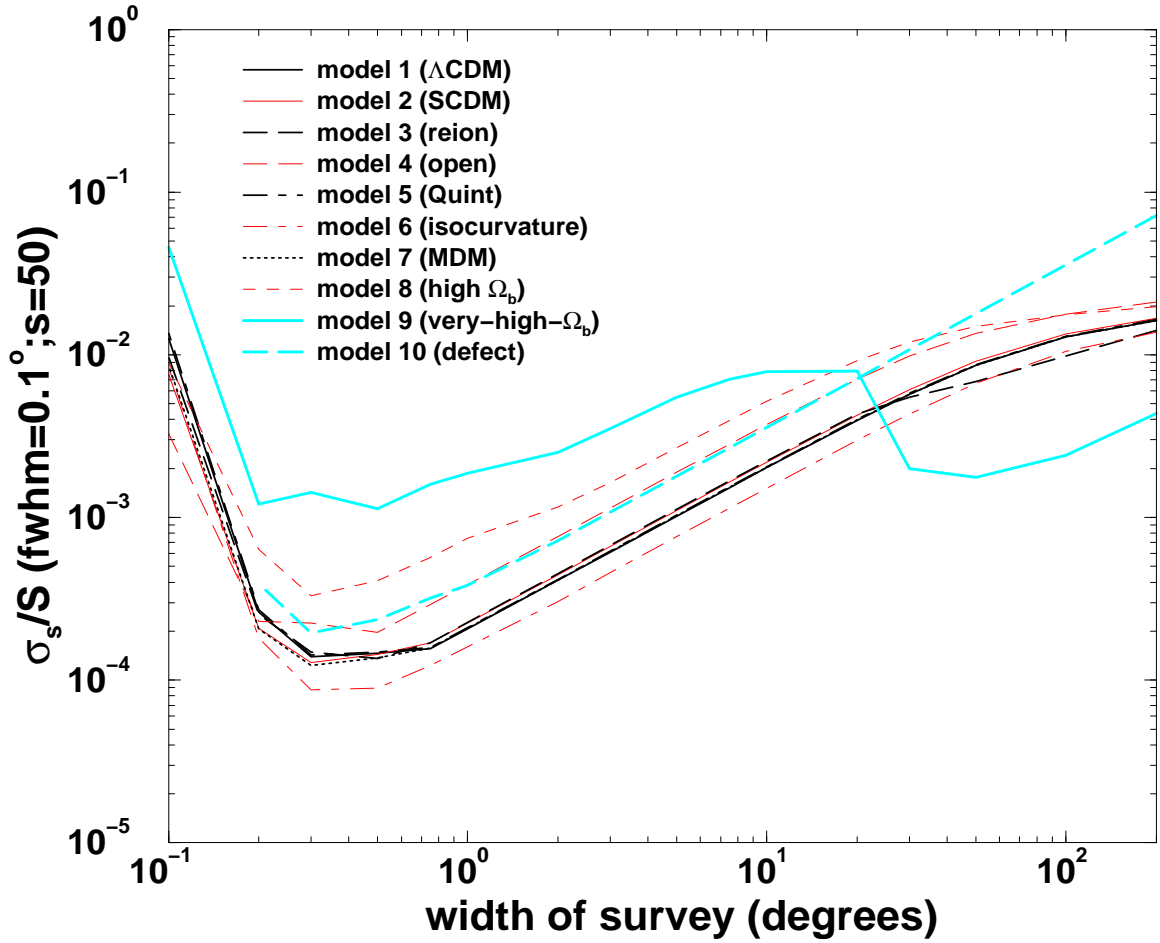,width=6in}}
\bigskip
\caption{The detectability of the polarization (as plotted in
         Fig.~\ref{fig:scalarsens}) for each of the models listed 
         in Table \ref{table}.}
\label{fig:sensitivities}
\end{figure}

\begin{table*}
\begin{center}
\begin{tabular}{|c|c|c|c|c|c|c|c|c|}
& model & $\Omega_0$ & $\Omega_\Lambda$& $\Omega_\nu$& 
 $h$ & $n$ & $\Omega_b h^2$ &
 $\tau_r$  \\ 
 \hline  \hline
 1 & $\Lambda$CDM & 1   & 0.7  & 0   & 0.65 & 1.0  & 0.020 & 0   \\ \hline
 2 & SCDM &1   & 0    & 0   & 0.40 & 1.0  & 0.019 & 0   \\ \hline
 3 & reion& 1   & 0.7  & 0   & 0.50 & 1.0  & 0.019 & 0.2 \\ \hline
 4 & open & 0.5 & 0    & 0   & 0.50 & 1.0  & 0.020 & 0   \\ \hline
 5 & Quint & 1 & 0.52   & 0   & 0.51 & 1.0  & 0.019 & 0   \\ \hline
 6 & isocurvature & 1   & 0    & 0   & 0.65 & 3.0 & 0.021 & 0   \\ \hline
 7 & MDM & 1   & 0    & 0.5 & 0.40 & 1.0  & 0.019 & 0   \\ \hline
 8 & high-$\Omega_b$ & 1   & 0    & 0   & 0.70 & 1.0  & 0.058 & 0   \\ \hline
 9 & very-high-$\Omega_b$ & 1   & 0    & 0   & 0.60 & 0.95 & 0.144 & 0.5 \\ \hline
10 & defect & 1 & 0 & 0 & 0.65 & N/A & 0.019 & 0 
\end{tabular}
\end{center}
\caption{Parameters for the models in
     Figs.~\protect\ref{fig:CTls}, \protect\ref{fig:ClGG}, and
     \protect\ref{fig:sensitivities}.  In model 5, the cosmological constant
     is actually a quintessence component with an equation of state
     $w=-0.5$. We get the power spectra for model 10 (the defect model) by
     scaling the power spectra of Ref. \protect\cite{SelPenTur97} so that the
     $C_\ell^{\rm TT}$ fit the data near the first peak.  Since
     Ref. \protect\cite{SelPenTur97} uses $\Omega_b h^2=0.0125$, we
     use the scaling discussed in the text \protect\cite{ZalHar95,husug} to
     approximate the polarization power spectrum for $\Omega_b h^2=0.02$.}
\label{table}
\end{table*}

To make this case, we have considered a number of models in
which the CMB power spectrum passes through recent data points
near $\ell\sim200$, as shown in Fig.~\ref{fig:CTls}.  The models
are listed in Table \ref{table}.  The $C_\ell^{\rm GG}$
polarization power spectrum for each model is shown in
Fig.~\ref{fig:ClGG}.  The detectability of the
polarization---{\it \'a la} Fig.~\ref{fig:scalarsens}---in
each of these models is shown in Fig.~\ref{fig:sensitivities}.
All of the models (except the high-baryon-density models) have
the BBN baryon-to-photon ratio, $\Omega_b h^2 = 0.02$.
Fig.~\ref{fig:ClGG} shows that all models with the BBN baryon
density produce roughly the same amount of polarization, and
Fig.~\ref{fig:sensitivities} shows more precisely that the
detectabilities are all similar.
The only models in which the
polarization signal is significantly smaller (and accordingly
harder to detect) are those with a baryon-to-photon ratio
that considerably exceeds the BBN value.

So why is this?  Heuristically, we expect the polarization amplitude to
be proportional to the temperature-anisotropy amplitude, and we have
fixed this.  This explanation is close, but still only partially
correct.  More accurately, the polarization comes from peculiar
velocities (the ``dipole'' \cite{husug}) at the surface of last scatter
\cite{ZalHar95}.  The peculiar-velocity amplitude is indeed proportional
to the density-perturbation amplitude that produces the peak in the
temperature power spectrum, but the constant of proportionality depends
on the baryon density \cite{ZalHar95,husug}; the peculiar velocity (and
thus the polarization) is larger for smaller $\Omega_b h^2$ (and the
dependence is actually considerably weaker than linear).  We also know
that the troughs in the temperature power spectrum are filled in by the
peculiar velocities.  Therefore, the polarization amplitude should
actually be proportional to the amplitude of the (yet-undetermined)
trough in the power spectrum, rather than the peak that has been
measured. Having fixed the peak height, the amplitude of the
trough---and thus the peculiar velocity---in turn depends only on the
baryon density, $\Omega_b h^2$.

In this way, the polarization amplitude depends primarily on the
baryon-to-photon ratio, itself proportional to $\Omega_b h^2$.  These
arguments further suggest that if the baryon density is significantly
higher than that allowed by BBN, then the polarization amplitude will be
smaller, and accordingly harder to detect.  There are many good reasons
to believe that the BBN prediction for the baryon density is robust.
However, it has also been pointed out that some problems (e.g., the baryon
fraction in clusters and a reported excess in power on $100\, h^{-1}$
Mpc scales) can be solved if we disregard the BBN constraint and
consider a much larger baryon density (seem e.g., Ref. \cite{eisetal}).
To illustrate, we also include in
Figs.~\ref{fig:CTls}, \ref{fig:ClGG}, and \ref{fig:sensitivities}
results from a high-baryon-density (i.e., $\Omega_b h^2=0.144$)
model \cite{eisetal}, and as expected, the polarization amplitude is
decreased relative to the temperature-fluctuation amplitude.  We
conclude that as long as the baryon density is not much larger than that
allowed by BBN, the results shown in Fig.~\ref{fig:scalarsens} should be
model-independent, but that the polarization may be significantly
smaller if the baryon-to-photon ratio is significantly higher.

\section{CONCLUSIONS}

We have carried out calculations that will help assess the prospects for
detection of polarization with various experiments.  Our results can be
used to forecast the signal-to-noise for the polarization signals
expected from density perturbations and from gravitational waves in an
experiment of given sky coverage, angular resolution, and instrumental
noise.  Even after the polarization has been detected, our results will
provide a useful set (although not unique) of figures of merit for
subsequent polarization experiments.  Of course, the ``theoretical''
factors considered here must be weighed in tandem with those that
involve foreground subtraction and experimental logistics in the design
or evaluation of any particular experiment. (As with
temperature-anisotropy experiments, these usually encourage
increasing the signal-to-noise to better distinguish systematic effects.)

In contrast to temperature anisotropies which show power on all scales
[i.e., $\ell(\ell+1)C_\ell\sim{\rm const}$], the polarization
power peaks strongly at higher $\ell$. Hence the signal-to-noise in a
polarization experiment of fixed flight time and instrumental
sensitivity may be improved by surveying a smaller region of sky, unlike
the case for temperature-anisotropy experiments.  The ideal survey for
detecting the curl component from gravitational waves is of order 2--5
degrees, and the sensitivity is not improved much for angular
resolutions smaller than 0.2 degrees.\footnote{Of course, better angular
  resolution and a larger survey area may be required to distinguish the
  gravitational-wave signal from a possible curl component from some
  foregrounds, nonlinear late-time effects \cite{lensing,vishniac},
  and/or instrumental artifacts.}  The polarization signal from density
perturbations is peaked at still smaller angular scales, and may be
better accessed by mapping an even smaller region of sky (again, keeping
in mind the caveats mentioned above).

Our numerical experiments and some physical arguments indicate
that the measured degree-scale temperature anisotropy fixes the
polarization amplitude in a model-independent way as long as we
use a fixed baryon density.  Thus, if the baryon density is
known from BBN, then {\it any experiment for which the curves in
Fig.~\ref{fig:scalarsens} fall below unity is guaranteed a
$3\sigma$ detection of the CMB polarization.}  A non-detection
would indicate unambiguously a baryon density significantly
higher than that predicted by BBN.

\acknowledgments

We thank L. Knox for providing the updated CMB data, and S. Hanany,
A.T. Lee and the whole MAXIMA team for inspiring some of the work
described herein.
This work was supported at Columbia by a DoE Outstanding Junior
Investigator Award, DEFG02-92-ER 40699, NASA NAG5-3091, and the Alfred
P. Sloan Foundation, and at Berkeley by NAG5-6552 and NSF KDI grant
9872979.


\begin{references}

\bibitem{rees} M. J. Rees, Astrophys. J. Lett. {\bf 153}, L1
     (1968).

\bibitem{arthur} A. Kosowsky, Ann. Phys. {\bf 246}, 49 (1996).

\bibitem{ZalHar95} M. Zaldarriaga and D. D. Harari, Phys.~Rev.~D 
     {\bf 52}, 3276 (1995).

\bibitem{Zal97} M. Zaldarriaga, Phys.~Rev.~D {\bf 55}, 1822 (1997).

\bibitem{Kos98} A. Kosowsky, astro-ph/9811163.

\bibitem{SpeZal97} M. Zaldarriaga and D. N. Spergel,
     Phys. Rev. Lett. {\bf 79}, 2180 (1997).

\bibitem{ourletter} M. Kamionkowski, A. Kosowsky, and
     A. Stebbins, Phys. Rev. Lett. {\bf 78}, 2058 (1997).

\bibitem{szletter} U. Seljak and M. Zaldarriaga,
     Phys. Rev. Lett. {\bf 78}, 2054 (1997).

\bibitem{KosLoe96} A. Kosowsky and A. Loeb, Astrophys. J. {\bf
     469}, 1 (1996).

\bibitem{HarHayZal96} D. D. Harari, J. Hayward, and
     M. Zaldarriaga, Phys.~Rev.~D {\bf 55}, 1841 (1996).

\bibitem{ScaFer97} E. S. Scannapieco and P. G. Ferreira,
     Phys.~Rev.~D {\bf 56}, 4578 (1997).

\bibitem{LueWanKam98} A. Lue, L. Wang, and M. Kamionkowski,
     Phys. Rev. Lett. {\bf 83}, 1503 (1999).

\bibitem{Lep98} N. Lepora,  gr-qc/9812077.

\bibitem{KamKos99} M. Kamionkowski and A. Kosowsky,
     astro-ph/9904108, Ann. Rev. Nucl. Part. Sci., in press (1999).

\bibitem{marcarthur} M. Kamionkowski and A. Kosowsky,
     Phys. Rev. D {\bf 57}, 685 (1998).

\bibitem{Lesetal99} J. Lesgourgues et al., gr-qc/9906098.

\bibitem{Kin98} W. H. Kinney, Phys.~Rev.~D {\bf 58}, 123506 (1998).

\bibitem{ZalSelSpe97} M. Zaldarriaga, U. Seljak, and
     D. N. Spergel, Astrophys. J. {\bf 488}, 1 (1997).

\bibitem{Lidetal97b} E. J. Copeland, I. J. Grivell, and
     A. R. Liddle, Mon. Not. R. Astron. Soc. {\bf 298}, 1233
     (1998).

\bibitem{MagHob97} J. Magueijo and M. P. Hobson, Phys. Rev. D {\bf
     56}, 1908 (1997).

\bibitem{HobMag96} M. P. Hobson and J. Magueijo,
     Mon. Not. R. Astron. Soc. {\bf 283}, 1133 (1996).

\bibitem{ourlongpaper} M. Kamionkowski, A. Kosowsky, and
     A. Stebbins, Phys. Rev. D {\bf 55}, 7368 
     (1997).

\bibitem{szlongpaper} M. Zaldarriaga and U. Seljak, Phys.~Rev.~D 
     {\bf 55}, 1830 (1997).

\bibitem{bjk98} J. R. Bond, A. H. Jaffe, and L. Knox, Phys. Rev. D {\bf
    57}, 2117, (1998).
  
\bibitem{bjk99} J. R. Bond, A. H. Jaffe, and L. Knox, Phys. Rev. D,
  submitted (1999), astro-ph/9808264.

\bibitem{jungmanone} G. Jungman, M. Kamionkowski, A. Kosowsky,
     and D. N. Spergel, Phys. Rev. Lett. {\bf 76}, 1007 (1996).

\bibitem{jungmantwo} G. Jungman, M. Kamionkowski, A. Kosowsky,
     and D. N. Spergel, Phys. Rev. D {\bf 54}, 1332 (1996).

\bibitem{ZibScoWhi99} J. P. Zibin, D. Scott, and M. White,
     astro-ph/9901028.

\bibitem{SelPenTur97} U. Seljak, U.-L. Pen, and N. Turok, Phys. Rev. Lett.
{\bf 79}, 1615 (1997).

\bibitem{husug} W. Hu and N. Sugiyama, Astrophys. J. {\bf 444},
     489 (1995).

\bibitem{eisetal} D. Eisenstein et al., Astrophys. J. Lett. {\bf
     494}, L1 (1998).

\bibitem{lensing} M. Zaldarriaga and U. Seljak, Phys. Rev. D {\bf 
     58}, 023003 (1998).

\bibitem{vishniac} W. Hu, astro-ph/9907103.

\end{references}
\end{document}